\documentclass[twocolumn,showpacs,preprintnumbers,amsmath,amssymb]{revtex4}

\usepackage{graphicx}
\usepackage{dcolumn}
\usepackage{bm}

\begin{document}

\preprint{APS/123-QED}

\title{Observation of the growth of a magnetic vortex in the transition layer of a mildly relativistic oblique 
plasma shock}

\author{G. C. Murphy\textsuperscript{1}, M. E. Dieckmann\textsuperscript{2}, L. O'C. 
Drury\textsuperscript{1}}

\affiliation{\textsuperscript{1} Dublin Institute for Advanced Studies, 31 Fitzwilliam
Place, Dublin 2, Ireland \\
\textsuperscript{2} VITA group, Department of Science and Technology (ITN), Link\"oping
University, Campus Norrk\"oping, SE-601 74 Norrk\"oping, Sweden}

\date{\today}

\begin{abstract}
A 2D particle simulation models the collision of two electron-ion plasma clouds along a quasi-parallel magnetic 
field. The collision speed is 0.9c and the density ratio 10. A current sheet forms at the front of the dense 
cloud, in which the electrons and the magnetic field reach energy equipartition with the ions. A structure 
composed of a solenoidal and a toroidal magnetic field grows in this sheet. It resembles that in the 
cross-section of the torus of a force-free spheromak, which may provide the coherent magnetic fields in 
gamma-ray burst (GRB) jets needed for their prompt emissions.
\end{abstract}

\pacs{52.35.Tc, 52.35.We, 52.65.Rr}
\maketitle

The ultrarelativistic jet of a Gamma Ray burst (GRB) gives rise to powerful eruptions 
of electromagnetic radiation, which can be detected across cosmological distances.
The non-stationary plasma acceleration at the source of the GRB jet results in a 
nonuniform plasma density and flow speed within the jet. Plasma clouds will thus 
collide at a mildly relativistic speed, triggering the formation of radiative shocks 
at the boundaries between individual clouds. The ensemble of these shocks is, according 
to the internal shock model \cite{ReMes,Piran}, the source of the ``prompt'' GRB emissions. 

Plasma processes and instabilities in the shock transition layer provide the relativistic 
electrons and the magnetic fields, which result in the emission of the electromagnetic 
radiation \cite{Filament,Bret,Schlickeiser}. The filamentation instability and the formation 
of relativistic shocks has been widely investigated numerically with the help of particle-in-cell 
(PIC) simulations for electron-positron plasmas \cite{Kazimura,Silva,Jaroschek} and for electron-ion 
plasmas \cite{Frederiksen,Spitkovsky,Martins}, which were all initially unmagnetized. However, the 
magnetic fields driven by the filamentation instability in an initially unmagnetized plasma tend to 
have too short a lifetime , and too small a coherence length to explain the prompt 
GRB emissions \cite{Waxman}.

The linear polarization of the GRB emissions suggests the presence of large-scale magnetic fields 
within the jets \cite{Lyutikov}. This serves as a motivation to examine further with PIC simulations 
the shock formation and evolution in the presence of an ambient magnetic field. Flow-aligned 
\cite{Hededal} and oblique \cite{Ohsawa,Dieckmann,Sironi} guiding magnetic fields have been introduced 
in previous PIC simulations of electron-positron and electron-ion shocks. In particular the oblique 
electron-ion shocks are formidable amplifiers of the magnetic fields and accelerators of the electrons. 
However, the studies in the Refs. \cite{Ohsawa,Dieckmann}, that could be representative for a magnetized internal 
GRB shock, employed a one-dimensional simulation geometry, which suppresses the beam 
filamentation and, as we show here, the formation of current vortices. 

\begin{figure}
\includegraphics[width=8.5cm]{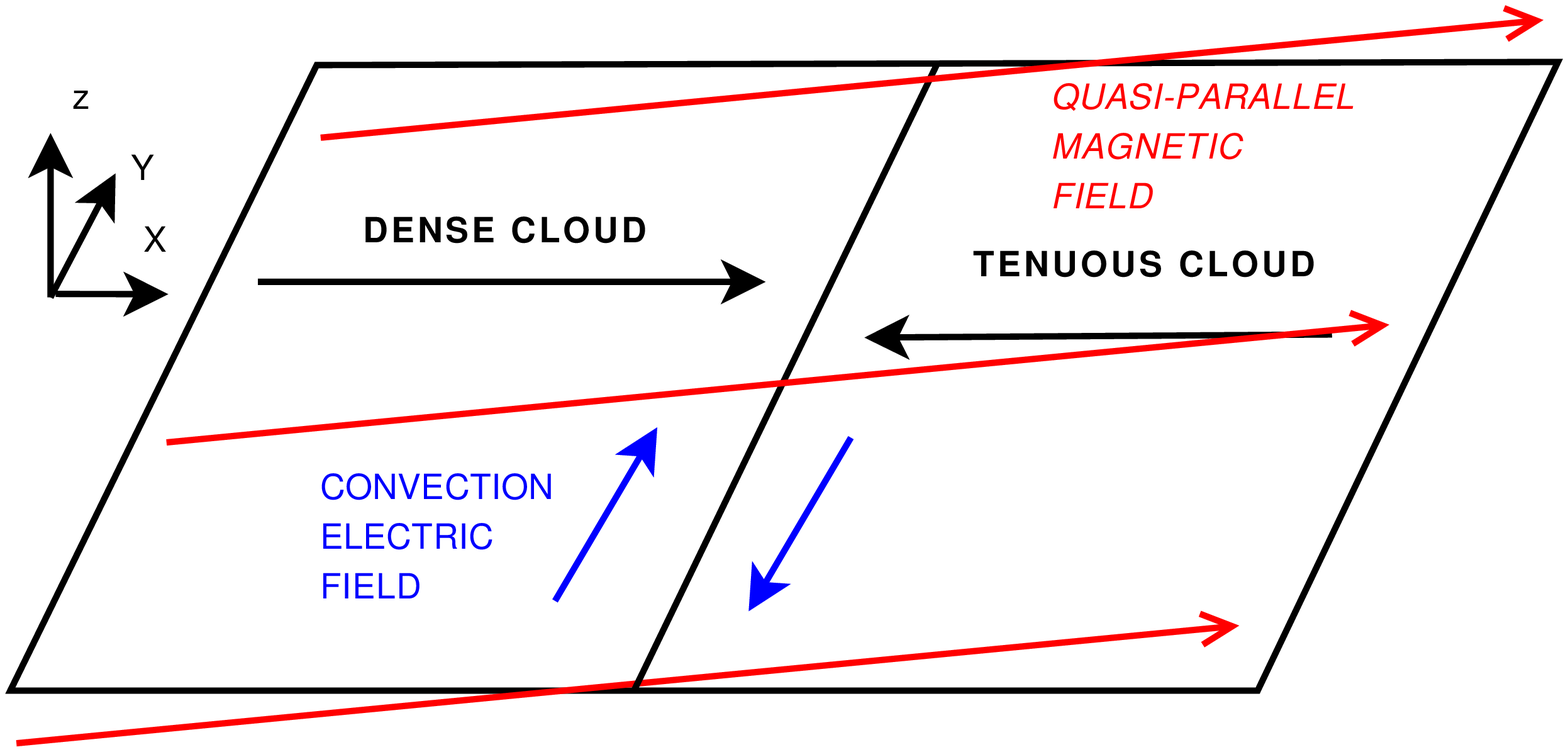}
\caption{Initial conditions: The simulation box is subdivided into two halves, one for each cloud. 
The clouds collide with the speed $v_c = 0.9c$ along $x$. The magnetic field ${\bf B}_0$ is oriented 
in the $x-z$ plane. The convection electric field ${\bf E}_0$ points along $y$ and changes its sign
at $x=0$.\label{fig1}}
\end{figure}

A PIC simulation code solves the Maxwell's equations and the relativistic Lorentz force equation 
for the computational particles \cite{Dawson}. We perform here a two-dimensional simulation study
with the widely used PSC code \cite{PSC}. Figure \ref{fig1} shows the simulation setup. 
Two plasma clouds, each consisting of ions and electrons with the mass ratio $m_i / m_e = 250$, 
collide at the position $x=0$. The initial electron and ion number densities of the dense cloud 
are both equal to $n_1$ and those of the tenuous cloud are both equal to $n_2 = n_1 / 10$. The 
velocity vectors of both clouds are antiparallel and aligned with $x$. The modulus $v_b$ of each 
cloud in the box frame gives the collision speed $v_c = 2v_b / (1+v_b^2/c^2) = 0.9c$. The dense 
cloud propagates to increasing values of $x$. The temperature of all 4 plasma species is 131 keV,
which yields an electron thermal speed $v_t = {(k_B T/m_e)}^{1/2}$, which is close to $v_b$. The 
cloud density ratio and the collision speed are both compatible with the internal shock model 
\cite{Piran}. 

The plasma frequency $\omega_{p2}={(n_1 e^2 / m_i \epsilon_0)}^{1/2}$ of the ions of the dense cloud 
normalizes all quantities, which have the subscript $p$ for MKS units. The time $t=\omega_{p2}\, 
t_p$, the position $(x,y)=(x_p,y_p) / \lambda_i$ with $\lambda_i = c / \omega_{p2}$, the velocity 
${\bf v}={\bf v}_p / c$, the electric and magnetic fields ${\bf E} = e {\bf E}_p / (m_i c \, \omega_{p2})$ 
and ${\bf B} = e {\bf B}_p / (m_i \omega_{p2})$. The momenta ${\bf p} = (p_x,p_y,p_z)$ 
are normalized as ${\bf p}= {\bf p}_p / m_i c$ (ions) and ${\bf p}= {\bf p}_p / m_e c$ (electrons). 
The initial ${\bf B}_{0p}$ gives a $\omega_{ce}= e|{\bf B}_{0p}|/m_e$ that equals 
the electron plasma frequency $\omega_{p1} = {250}^{1/2} \omega_{p2}$ of the dense cloud. We obtain 
$|{\bf B}_0|=250^{1/2}$ with $B_{0x}=10B_{0z}$. The modulus of the convection electric field is 
$|E_{0y}| = v_b B_{0z}/c$. The simulation box size $L_x \times L_y = 656 \lambda_i \times 6 \lambda_i$ 
is resolved by $2.8 \cdot 10^4 \times 256$ rectangular grid cells. The dense (tenuous) plasma is resolved 
by 100 (50) particles per cell per species. We employ periodic boundary conditions in all directions and 
no particles are introduced into the simulation after $t=0$. The plasma clouds detach instantly from the 
boundary at $x=\pm L_x/2$ and they will interpenetrate at $x=0$. We will discuss in what follows the time evolution as shown in the movies followed by the 
plasma state at the simulation's end at the time $t=180$.

In the movies we show the whole time integrated evolution of the relevant parameters from $t=1 \omega_{p,ion}^{-1}$ to $t=180 \omega_{p,ion}^{-1}$.
Movie 1 shows the separate evolution of the field components $B_x, B_y, B_z$.
Movie 2 shows the evolution of $J_z$ in the upper panel and $\log |J_x + i J_y|^2$ in the lower panel.
In Movie 1 we see at early times the field is planar.
The fields have a dipolar structure.
As time elapses it becomes more and more nonplanar, filamented and fragmentary.
The $B_x$ which is initially uniform assumes a fragmented morphology, with gradually increasing structural lengthscales.
At early times the forming shock accelerates before reaching a steady speed.
Striped magnetic fields - tell-tale signs of the filamentation instability in the $B_x$ and $B_z$ movies before $t < 43 \omega_{p,ion}^{-1}$.
Gradually the magnetic fields become dominated by large-scale structures which represent the extrema.
$B_z$, a tracer of the plane current takes on the almost circular structure characteristic of the ring current.
In Movie 2 we see the gradual breakup of the initial current sheet and the formation of a current ring.
The sheet initially breaks up into smaller rings at $t \sim 82  \omega_{p,ion}^{-1}$ which gradually increase in volume and merge ($t \sim 113  \omega_{p,ion}^{-1}$) to form the final large rings, limited only by the box size.
Incoming current filaments are deflected from the current sheet and shear off in the negative y direction, rolling up into vortex structures which later merge and distort the current sheet.

\begin{figure}
\includegraphics[width=8.5cm]{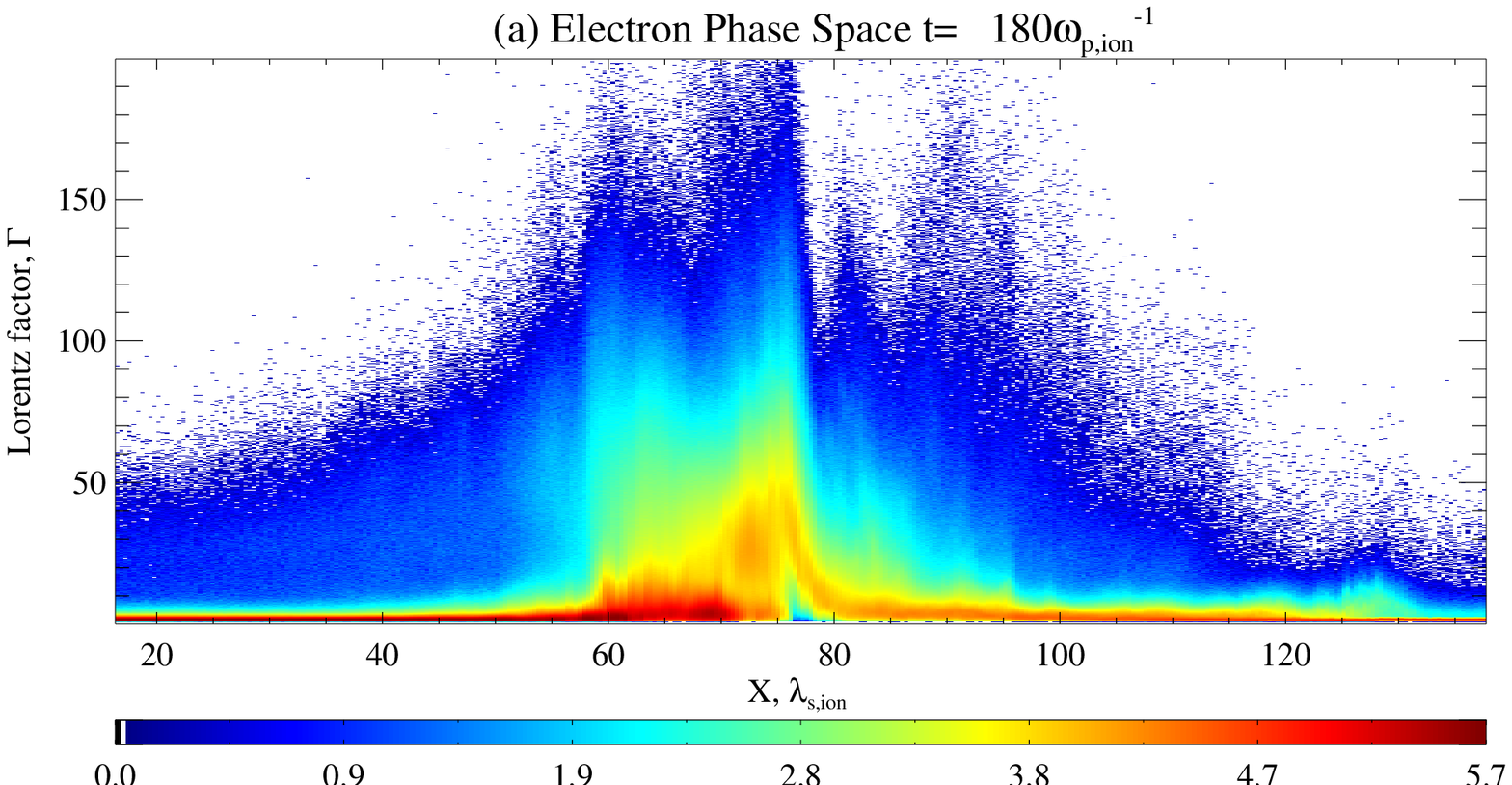}
\includegraphics[width=8.5cm]{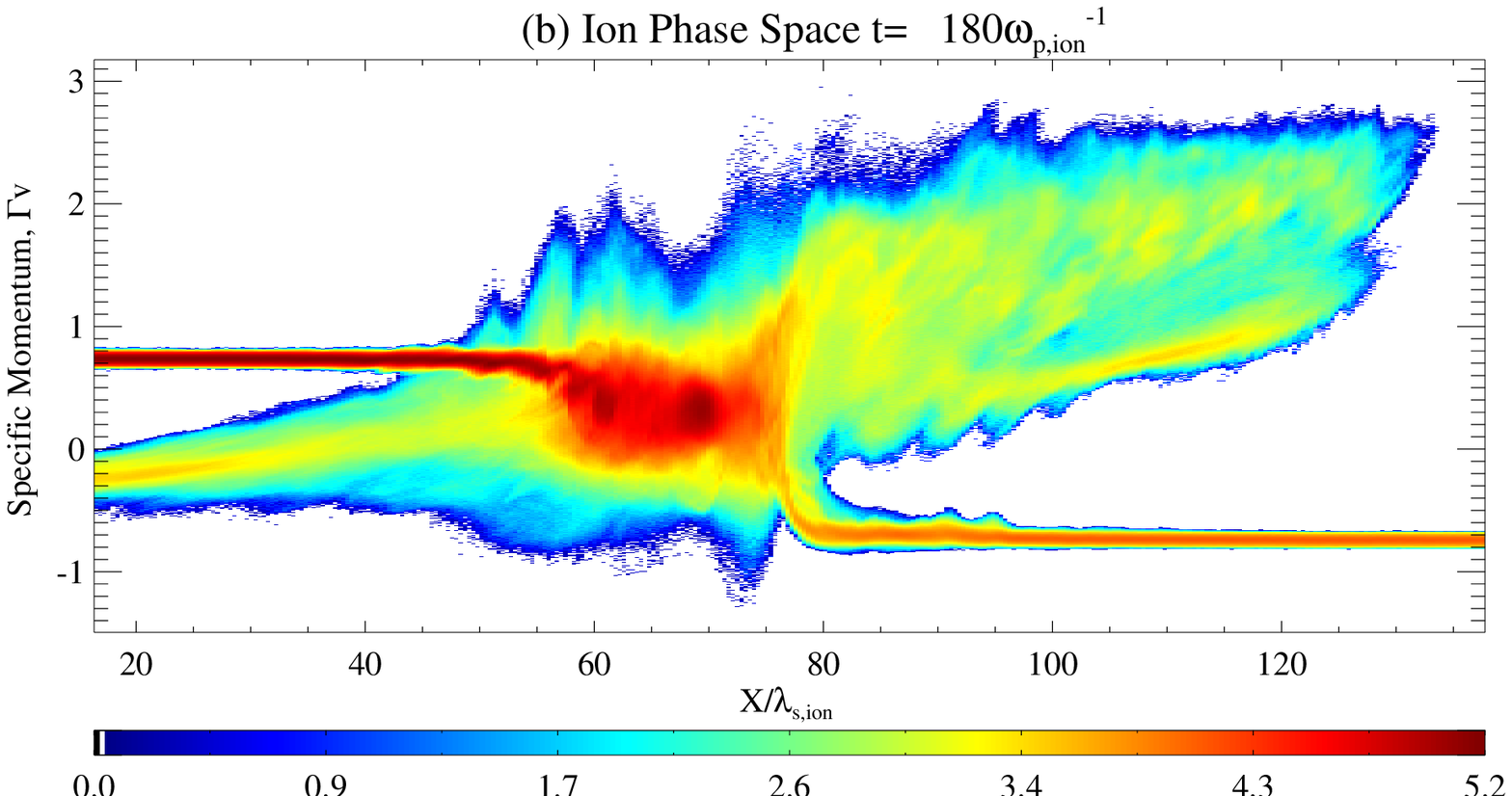}
\caption{The electron phase space density in the $x-\Gamma$ plane (a) and the ion phase space density 
in the $x-p_x$ plane (b): The $\Gamma > 200$ reached by some electrons implies that their kinetic energy 
is comparable to that of the ions. The ion distribution reveals a forming downstream region $55<x<70$
that separates a reverse shock from the forward shock, which moves to increasing $x$. The ions of the
tenuous cloud are reflected back into the upstream region within $70<x<80$. This interval coincides 
with that of the strongest electron acceleration. The color scale is 10-logarithmic.\label{fig2}}
\end{figure}

Figure \ref{fig2} displays the phase space distributions $f_e (x,\Gamma)$ of the electrons and 
$f_i (x,p_x)$ of the ions, which have been integrated over all $y$. The $f_e$ shows us the energies 
the electrons reach at the shock, while the merging of the ions of both clouds in $f_i (x,p_x)$ is 
a necessary condition for a plasma shock. 
A downstream region is visible in $f_i$ for $55<x<70$, which separates a reverse shock from the forward 
shock moving to increasing $x$. The reverse shock is weaker than the forward one, due to the asymmetric
plasma densities that yield a net flow of the downstream region. In what follows we discuss only the 
forward shock. Electrons reach in the cloud overlap layer values of $\Gamma \approx 100$ in significant 
numbers and the fastest ones reach $\Gamma \approx 250$. An ion with the mass $m_i =250$ that moves with 
$v_b$ has in the simulation frame the kinetic energy of an electron with $\Gamma \approx 70$. A shock 
transition layer with $70<x<80$ reflects the incoming ions of the tenuous cloud and accelerates its 
electrons, as observed at the Earth's perpendicular bow shock \cite{Anderson}.

\begin{figure}
\includegraphics[width=8.5cm]{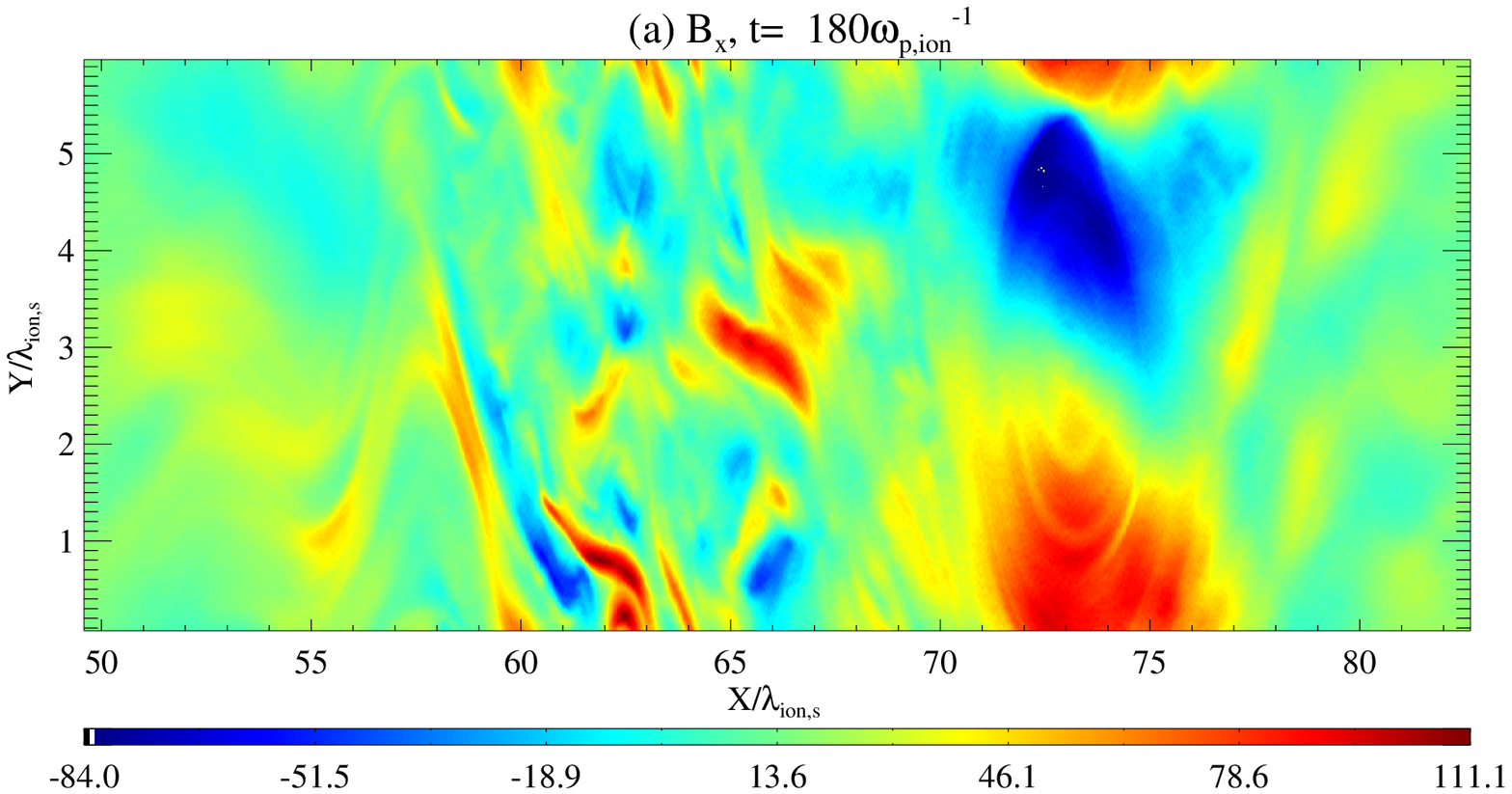}
\includegraphics[width=8.5cm]{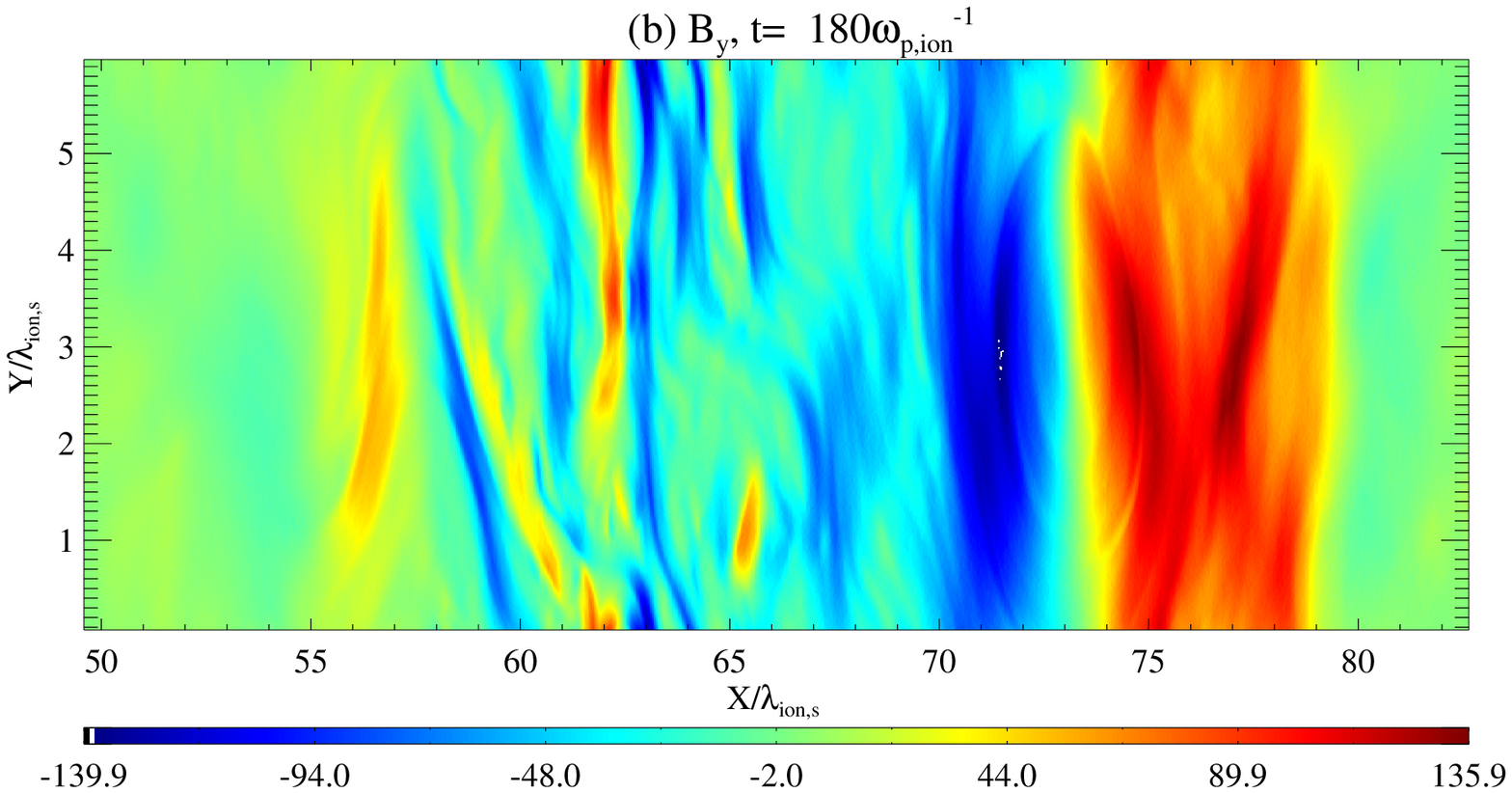}
\includegraphics[width=8.5cm]{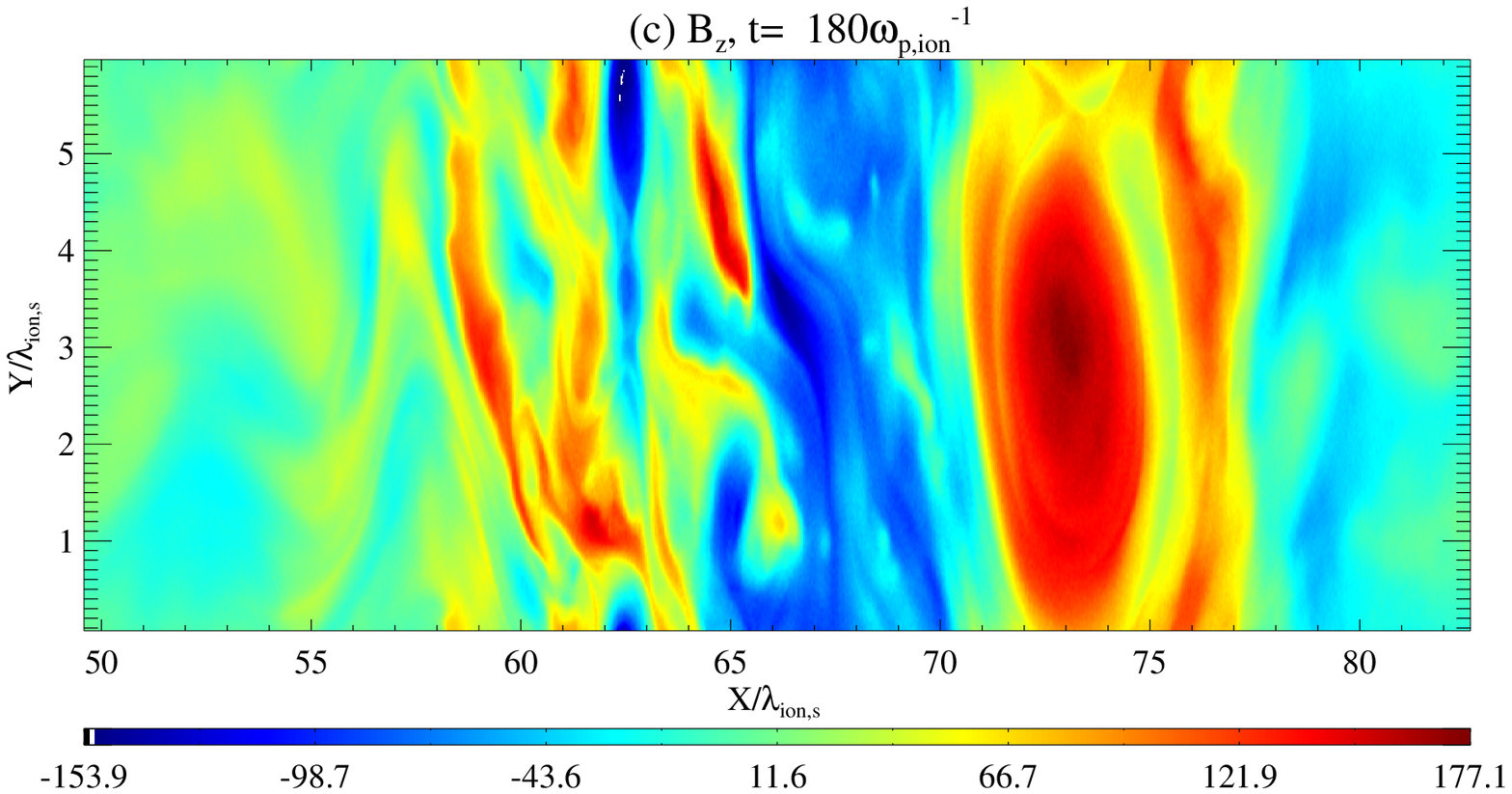}
\caption{The magnetic field components $B_x$ (a), $B_y$ (b) and $B_z$ (c): All components evidence 
structures on ion skin depth scales within the $55<x<70$. The $B_x$ and $B_y$ reveal a magnetic vortex
with a counterclockwise sense of rotation within $70<x<77$ and with its center with $|B_x + iB_y| 
\approx 0$ at $(x,y) \approx (74,3)$. A further quasi-planar structure in $B_y$ is located within 
$77<x<80$. The $B_z$ is strongest in the interior of the vortex in the $x-y$ plane and its perimeter 
coincides with the magnetic vortex.\label{fig3}}
\end{figure}

Figure \ref{fig3} depicts ${\bf B}$, which has grown in amplitude by an order of magnitude compared
to $|{\bf B}_0| \approx 16$.
All magnetic components reveal elongated structures in the downstream region $55<x<70$, which have
a thickness of up to one ion skin depth. This magnetic field is tied to the current channels driven 
by the filamentation instability and the ${\bf B}_0$ deflects the particles and thus the currents 
in all directions, causing them to form three-dimensional distributions. A massive coherent magnetic 
structure is present in the shock transition layer with $70<x<80$. The $B_x$ and $B_y$ form a vortex, 
which rotates counterclockwise in the $x-y$ plane and is centered at $(x,y) \approx (74,3)$, where 
its modulus $|B_x + iB_y| \approx 0$. The vortex fills up the full $y$-interval of the simulation box 
at this time. The $B_z$ has its peak at the center of the magnetic vortex and its amplitude decreases, 
radially from the centre $(x,y) \approx (74,3)$.

\begin{figure}
\includegraphics[width=8.5cm]{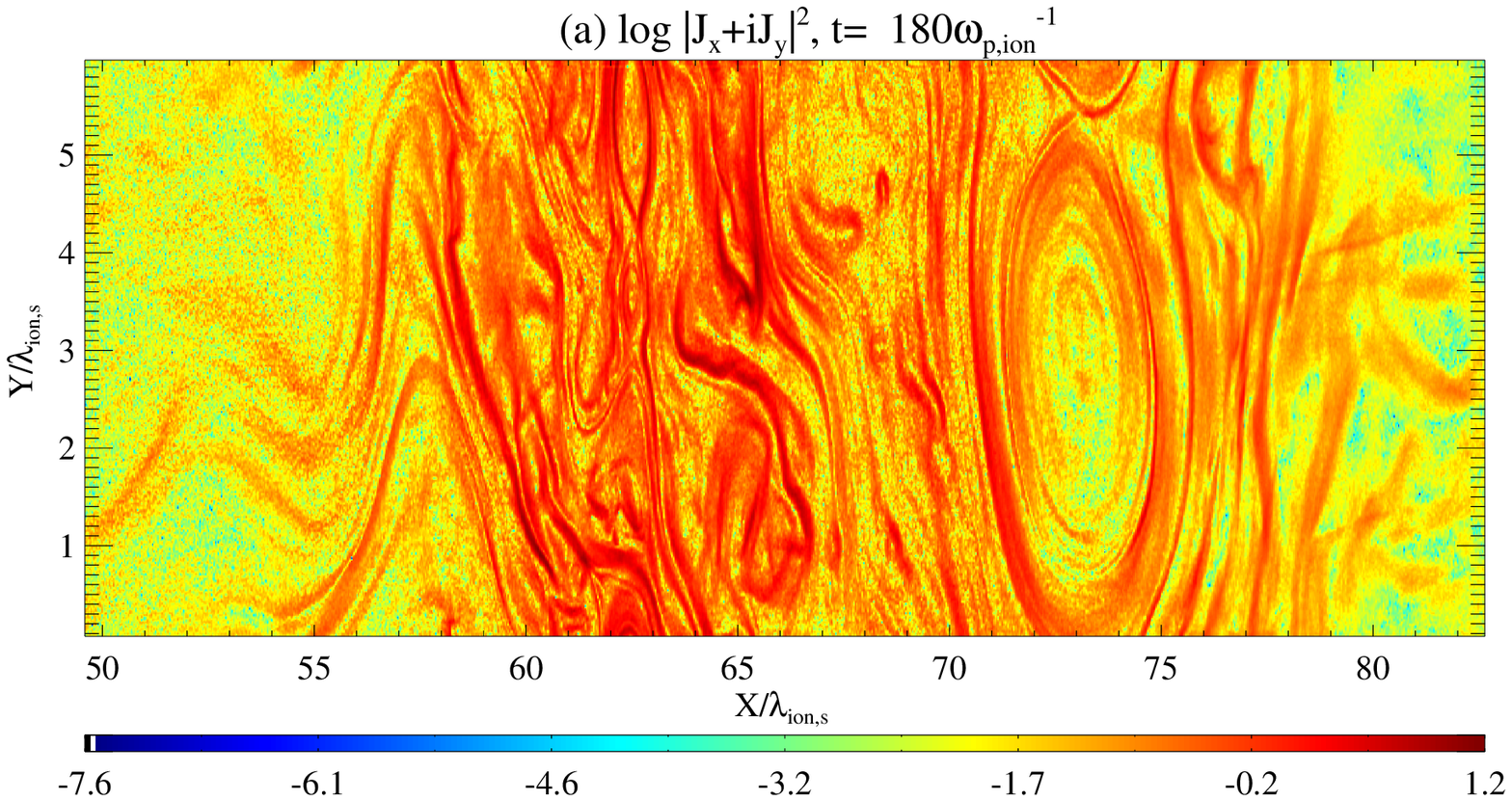}
\includegraphics[width=8.5cm]{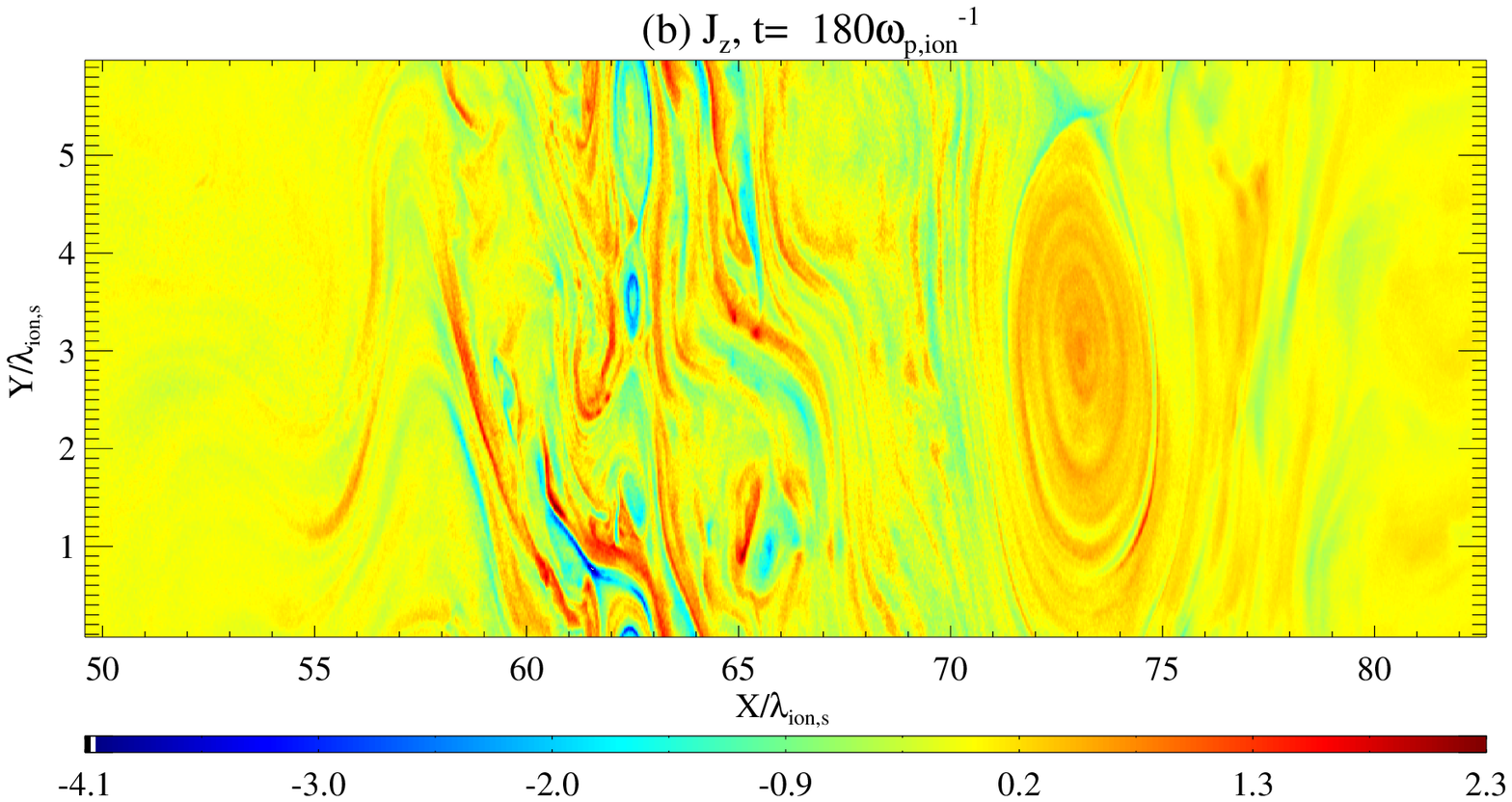}
\caption{The currents $\log_{10} (J_x^2 + J_y^2)$ (a) and $J_z$ (b): The distribution in (a) reveals 
an almost circular vortex for $71<x<76$ and $-0.5 < y < 5$ (periodic wrap-around). This vortex 
encircles an interval with a strong $J_z>0$. The currents show a correlation indicating that it flows 
in the plane on the perimeter but is increasingly deflected along $z$ as we go to the center. A 
current system is present in all components in the downstream region $55<x<70$ and in $76<x<80$.
\label{fig4}}
\end{figure}

The magnetic field shows a clear subdivision between its components in the simulation plane and the 
component orthogonal to it, which should hold also for the currents. Figure \ref{fig4} shows the
currents in and out of the simulation plane.
In agreement with the magnetic structure in Fig. \ref{fig3}, the currents in Fig. \ref{fig4} are
filamentary in the forming downstream region $55<x<70$ and in the interval $76<x<80$, while they 
look fundamentally different in the shock transition layer in $70<x<76$. Here $J_x$ and $J_y$ form 
a vortex centered at $(x,y) \approx (74,3)$, while $J_z>0$ is strong in the ellipsoidal interval 
encircled by the vortex. Current channels almost parallel to $y$ are located within $76<x<80$, 
which give rise to the quasi-planar $B_y$ at this position in Fig. \ref{fig3}(b). These current
striations are what is left of the current sheet, which developed due to the different magnetic 
deflection of the upstream electrons and ions by the perpendicular magnetic field component of 
the shock.

Figures \ref{fig3} and \ref{fig4} reveal why the magnetic field structure in the shock transition 
layer appears to be stable. The current vortex in the $x-y$ plane is responsible for $B_z$. This 
current is almost aligned with the magnetic field in the simulation plane in a region with 
$B_z \approx 0$, as Figs. \ref{fig3}(b) and \ref{fig4}(a) show. The current vortex thus experiences 
negligible magnetic force. Its sense of rotation is counter-clockwise and it gives an interior 
magnetic field that points in the positive $z$-direction. The $B_z$ outside the current vortex is 
negligible and this system is analogous to an infinitely long coil with a symmetry axis parallel 
to $z$. The current $J_z$, which is responsible for the vortex formed by $B_x$ and $B_y$, also flows 
along ${\bf B}$, because $|B_x + iB_y| \approx 0$ in the centre of the magnetic vortex. However,
the magnetic field configuration is probably only approximately force-free, because the current 
vortex in Fig. \ref{fig4}(a) spirals into the interval with $B_z \neq 0$ thereby coupling the
current components. A force-free magnetic field must fullfill $\nabla \times {\bf B} = a{\bf B}$ 
with a constant $a$ \cite{Old,New}, which is achieved here by the combination of a solenoidal 
component and of a ring component of ${\bf B}$ with a comparable strength.

Discussion: Our simulation demonstrates the growth of the filamentation instability, in spite of
the initial conditions that reduce its growth rate relative to the competing instabilities
\cite{Dieckmann}. The filamentation instability is, however, not the main source of the magnetic 
field within the shock transition layer. A current sheet develops at the front of the dense cloud,
because the electrons and ions are not equally deflected by the perpendicular component of ${\bf B}$. 
This current sheet boosts the component of ${\bf B}$ orthogonal to the flow direction to an amplitude
$\approx 100 B_{0z}$, resulting in a magnetic energy density that is comparable to the initial ion 
kinetic energy density. This current sheet is unstable and a magnetic structure develops, which is 
composed of solenoidal and ring components. Such a topology can, in principle, be force-free. However, 
the current loop in the simulation plane is not closed and it is also elliptical. In fact, we cannot 
expect this magnetic field structure to be force-free. The ram pressure of the inflowing upstream 
plasma, the electron acceleration to extreme speeds and the ion reflection in the shock 
transition layer exert a force on this structure and they are probably responsible for its compression 
along $x$ into an ellipse. The spiral structure of the current in the simulation plane may be a remnant 
of the growth of the structure and thus a transient effect.   

Clearly our simulation results are affected by the periodic boundary conditions in the $y$ direction 
and by the spatial 2D geometry. The 2D geometry assumes that the magnetic structure is an infinitely
long cylinder with the symmetry axis along $z$. A force-free 3D topology is obtained if the cylinder 
is bent to form a closed loop known as a spheromak, which is composed of poloidal and toroidal magnetic 
fields \cite{Old,New}. The structure we observe in our simulation would then be a cross-section of the 
torus. The obvious effect of the periodic boundary conditions along $y$ is to halt the growth of the 
magnetic structure, by which a steady state is reached (see the movies). This steady state is actually quite remarkable, 
considering the powerful particle acceleration processes taking place at this location. A wider $y$ 
interval will presumably result in a further growth of the magnetic structure. Spheromaks can be
quite large and they have been invoked to explain large scale structures in the solar wind triggered by 
coronal mass ejections \cite{Kataoka}. The poloidal magnetic field of the spheromak, which corresponds 
to $B_z$ in our simulation, is strong and homogeneous within the torus' perimeter. 

Our simulations suggest that spheromaks, or similar magnetic loops, that form within the current sheet 
of an internal GRB shock may thus provide the means to grow stable structures with a strong coherent 
magnetic field from kinetic scales to MHD scales. The current sheet, which is the source of this structure,
does require an ambient magnetic field with a component orthogonal to the flow velocity vector and ions.

Acknowledgments: This work was supported by Science Foundation Ireland (SFI) grant number 08/RFP/PHY1694.
Dr ME Dieckmann acknowledges support by Vetenskapsr\aa det. 
The SFI/HEA Irish Centre for High-End Computing (ICHEC) provided 
computational facilities and support.

\end{document}